\documentclass[conference]{IEEEtran}

\usepackage[utf8]{inputenc}
\usepackage[T1]{fontenc}
\usepackage{hyperref}
\usepackage{graphicx}
\usepackage{subcaption}
\usepackage{comment}
\usepackage{multirow}
\usepackage[table]{xcolor}
\usepackage{xcolor}
\usepackage[]{algorithm2e}
\usepackage{url}
\usepackage{soul}
\usepackage[hang,flushmargin]{footmisc}
\newcommand{\furl}[1]{\footnote{\scriptsize \url{#1}}}
\colorlet{punct}{red!60!black}
\definecolor{background}{HTML}{EEEEEE}
\definecolor{delim}{RGB}{20,105,176}
\colorlet{numb}{magenta!60!black}

\title{\LARGE \bf Quality Prediction of Open Educational Resources
\\ \Large A Metadata-based Approach}

\author{
    \IEEEauthorblockN{Mohammadreza Tavakoli$^{1}$, Mirette Elias$^{2}$, G\'abor Kismih\'ok$^{1}$, S\"oren Auer$^{1}$}
    \IEEEauthorblockA{$^{1}$TIB – Leibniz Information Centre for Science and Technology, Germany \{reza.tavakoli, gabor.kismihok, soeren.auer\}@tib.eu
    \IEEEauthorblockA{$^{2}$University of Bonn and Fraunhofer IAIS, Germany \{melias@uni-bonn.de\}}
    }
}
\begin{document}

\maketitle

\begin{abstract}
In the recent decade, online learning environments have accumulated millions of Open Educational Resources (\emph{OERs}). However, for learners, finding relevant and high quality OERs is a complicated and time-consuming activity. Furthermore, metadata play a key role in offering high quality services such as recommendation and search. Metadata can also be used for automatic OER quality control as, in the light of the continuously increasing number of OERs, manual quality control is getting more and more difficult. 

In this work, we collected the metadata of 8,887 OERs to perform an exploratory data analysis to observe the effect of quality control on metadata quality. Subsequently, we propose an OER metadata scoring model, and build a metadata-based prediction model to anticipate the quality of OERs. Based on our data and model, we were able to detect high-quality OERs with the \emph{F1 score} of \textbf{94.6\%}.
\end{abstract}


\section{Introduction}

Nowadays, Open Educational Resources (OERs) play a significant role in informal learning. However, the low quality of search services and recommender systems limit the use of OERs~\cite{chicaiza2017recommendation}. Obviously, the lack of metadata that thoroughly describe OERs has a negative effect on the performance of these services~\cite{kiraly2018measuring,ochoa2009automatic}.

Furthermore, the vast amount of OERs, which are provided daily by content creators around the world, forces us to put more emphasis on automatic controlling of OERs quality. We depart from a strong assumption that OERs metadata and content quality have tight relationship with each other: the more OERs have high quality metadata, the higher the probability of high quality content is. Currently, manual methods are often used to evaluate both the quality of OER content and metadata~\cite{Tani2013}, which solutions are time consuming and not scalable ~\cite{ochoa2009automatic}. Therefore, we expect that a thorough automatic metadata analysis will significantly improve the quality control of OERs. 
There are some attempts already, which aim at automatizing the quality assessment of metadata (e.g., \cite{ochoa2009automatic,Trippel2014,nichols2008lightweight}).
However, these  focus only on the criteria definitions and metrics to evaluate existing OER metadata~(e.g., \cite{bruce2004continuum,Ochoa2006,romero2019proposal}) without building an intelligent model, or models, to predict the quality of OERs based on metadata.

In this paper, we discuss the details of our exploratory data analysis on the metadata of 8,887 OERs from SkillsCommons\footnote{\url{http://skillscommons.org}}, in order to provide insights about the quality of metadata in existing OERs, and the effect of quality control on metadata quality. Then, we build a metadata-based scoring and prediction models to anticipate the quality of OERs based on the results of our exploratory analysis.


\section{Related Work}\label{sec-related} 
OER metadata is important not only to aid learners in finding relevant content among large amount of OERs, but also to indicate OER quality~\cite{ushakova_2015}. 
In the literature, the quality of OER metadata has been determined in terms of the following dimensions: completeness, accuracy, provenance, conformance to expectations, logical consistency and coherence, timeliness, and accessibility~\cite{bruce2004continuum}. 
Ochoa and Duval have converted these dimensions into a set of calculated metrics, which have been reused by most of the researchers addressing quality of OER metadata~\cite{Ochoa2006}.
They partially evaluated their metrics (i.e., completeness, accuracy) on a list of 425 OERs from the ARIADNE Learning Object Repository~\cite{ochoa2009automatic}.

Most studies about OER metadata quality have mainly focused on the completeness of metadata, by means of the availability of metadata elements, the presence of their values~\cite{sicilia2005complete}, and the evaluation of those values~\cite{margaritopoulos2012quantifying}. 
Pelaez and Alarcon have evaluated the completeness and consistency of OERs \cite{pelaez2017metadata} by building their calculation on Ochoa and Duval's metrics \cite{Ochoa2006}. They evaluated consistency of metadata elements values with respect to the standardized domain values (e.g. Language should be according to ISO 639-111 language standard).
However, most of these approaches are not automatic, and either conceptual \cite{margaritopoulos2008conceptual}, \cite{romero2019proposal} or focusing on one, or only a few dimensions \cite{margaritopoulos2012quantifying}, \cite{romero2018exploring}. Therefore, there is a need for automatic and intelligent metadata quality assessment in order to improve the discoverability, usability, and reusability of OERs \cite{gavrilis2015measuring}.

Based on the state-of-the-art, it is clear that: 1) it is worthwhile and timely to analyze OER metadata to improve OER-based services; and 2) there is a lack of intelligent prediction models, which evaluate the quality of OERs based on their metadata to facilitate the quality control. For these reasons the main research questions and objectives of our current work are:
\begin{itemize}
    \item Conducting an exploratory data analysis on large amount of OERs' metadata.
    \item Building a scoring model with a data-driven approach that helps OER repositories and authors to evaluate and improve the quality of their OER metadata.
    \item Predicting the quality of OERs based on their metadata. This should guide automatic quality control processes and ultimately result in higher OER quality.
\end{itemize}

\section{Data Collection and Research Method}\label{sec-method}
In this section, we explain our steps towards our proposed model. First, we collected and maintained a large dataset of OER metadata. Second, we performed an exploratory data analysis and deduced results. Third, we built a scoring model accordingly, and finally, we proposed a prediction model to anticipate the quality of OERs.

\subsection{Data Collection}
We built an OER metadata dataset after retrieving all search results for the terms "Information Technology" and "Health Care" via the \emph{SkillsCommons API}\footnote{\url{http://support.skillscommons.org/home/discover-reuse/skillscommons-apis/}} resulting in a metedata pool of 8,887 OERs\footnote{Our dataset can be downloaded from: \url{https://github.com/rezatavakoli/ICALT2020_metadata}}. 
Each OER contains the following metadata: url, title, description, educational type, date of availability, date of issuing, subject list, target audience-level, time required to finish, accessibilities, language list, and quality control (a categorical value that shows if a particular OER went through manual quality control or not).

\subsection{Exploratory Analysis of OER Metadata}\label{sec-analysis}
As a point of departure, we used our dataset to explore the availability of metadata values, which are related to the category quality control ("with control" or "without control"). Our analysis showed a clear increase in OER metadata quality (in terms of \emph{availability} of metadata) in the quality controlled OERs, which can be interpreted as a result of OER quality control. However, our analysis also indicated that the proportion of manual OER quality control in our dataset has been decreasing over the last years (from more than 60\% in 2016 to less than 40\% in 2019). We believe that the growing number of OERs is among the main reasons for this change. 
To conclude the results of our exploratory analysis:
\begin{enumerate}
    \item Quality controlled OERs can be used to define benchmarks for quality of metadata fields
    \item There is a need to define a method that facilitates the automatic assessment of OER metadata quality, and consequently the quality control of OERs. 
\end{enumerate}

\subsection{OER Metadata Scoring Model}
As the first step when building our scoring model, we defined the importance of each metadata field based on those OERs, which went through quality control.

For this purpose, we set the importance rate of each metadata field according to its availability rate among quality controlled OERs (between 0 and 1). For instance, all quality controlled OERs have a \emph{title} and therefore, we set the importance rate of \emph{title} to $1$, and for \emph{Time Required}, we set it to 0.58 since 58\% of the controlled OERs have \emph{Time Required} metadata. Moreover, we normalised the calculated importance rates as normalized importance rate.

Afterwards, for each field, we created a rating function in order to rate metadata values. We fit a normal distribution on values (lengths) of the following metadata fields: \emph{title, description, and subjects}, as they have distributions similar to normal and used the reverse of \emph{Z-score} concept (as $\frac{1}{\lceil |x-\bar{x}|/s\rceil}$ where $\bar{x}$ and $s$ is the mean and standard deviation respectively of the field in the dataset) to rate the metadata values based on the properties of the quality controlled OERs. Thus, the closer an OER \emph{title/description/subjects} length is to the mean of distributions, the higher is the rate. It should be mentioned that when a value is equal to the mean, the rate will be 1 and when it is empty the rate will be 0. Moreover, we used a boolean function for the four fields: \emph{level, length, language, and accessibility} which assigns 1 when they have a value and assigns 0 otherwise. Table~\ref{tab:tbl-metadata} illustrates the metadata fields, importance rates, normalized importance rates, and the rating functions.
\begin{table}[h]
\caption{OER metadata fields and importances}\label{tab:tbl-metadata}
\begin{tabular}{|p{1.7cm}|>{\centering\arraybackslash}p{1.3cm}|>{\centering\arraybackslash}p{1.5cm}|>{\centering\arraybackslash}p{2.6cm}|}
\hline
\rowcolor[HTML]{C0C0C0} 
\textbf{Type} & \textbf{Importance Rate [0-1]} & \textbf{Normalized Importance Rate [0-1]} & \textbf{Rating Function [0-1]} \\ \hline
Title & 1 & 0.17 & $\frac{1}{\lceil |x-5.5|/2.5\rceil}$ \\  \hline
Description & 1 & 0.17 & $\frac{1}{\lceil |x-54.5|/40\rceil}$ \\  \hline
Subjects & 0.86 & 0.145 & $\frac{1}{\lceil |x-4.5|/3.5\rceil}$ \\  \hline
Level & 0.98 & 0.165 & If available: 1; else: 0 \\  \hline
Language & 0.92 & 0.155 & If available: 1; else: 0 \\  \hline
Time Required & 0.58 & 0.098 & If available: 1; else: 0 \\  \hline
Accessibilities & 0.59 & 0.099 & If available: 1; else: 0 \\  \hline
\end{tabular}
\end{table}

Finally, we defined the following two scoring models in order to cover the availability and adherence of the defined benchmarks:

\textit{\textbf{Availability Model}}.
We calculate the availability score of an OER $o$ as Equation~(\ref{eq:availability}) where $norm\_import\_rate(k)$ is \emph{Normalized Importance Rate} of metadata field $k$. This score shows how complete that metadata is in a weighted summation, in which the normalized important rates are the weights. Therefore, the more an OER contains important fields, the higher the availability score is. For instance, an OER with metadata about \emph{title}, \emph{description} and \emph{level} (metadata fields with the highest importance rates), achieves a higher availability score than another one which has metadata for \emph{subjects}, \emph{time required}, and \emph{accessibilities}.
\begin{equation} \label{eq:availability}
  avail\_score(o) = \hspace{-3mm} \sum_{k=available fields} \hspace{-3mm} norm\_import\_rate(k)
\end{equation}

\textit{\textbf{Normal Model}}.
We calculate the normal score of an OER $o$ as Equation~(\ref{eq:adherence}), where $norm\_import\_rate(k)$ is the \emph{Normalized Importance Rate} of metadata field $k$, and \emph{rating(o, k)} is the assigned rating to OER $o$ based on the rating function of $k$. This score shows how close metadata to the defined benchmark is (based on metadata of the OERs with quality control). With this scoring model, an OER which has the most similar metadata properties with the metadata of quality controlled OERs, achieves the highest normal score.
\begin{equation} \label{eq:adherence}
  norm\_score(o) = \hspace{-3mm} \sum_{k=fields} \hspace{-3mm} norm\_import\_rate(k) * rating(o,k)
\end{equation}

\subsection{Predicting the quality of OERs based on their metadata} \label{sec-model}
We used 80\% of our data as a training set and trained a machine learning model to predict the quality of OERs based on their metadata and our scoring model. Therefore, we got the OERs “with control” as higher quality class (containing 4,651 OERs), and set the remaining as lower quality class (containing 4,236 OERs). As a classifier, a Random Forest model was trained to make a binary decision (i.e., high-quality or low-quality) based on the fields: \emph{Importance score, Availability Score, Level Metadata Availability, Description Length, Title Length, and Subjects Length}. 

\section{Validation} \label{sec-validation}
We built a test set using the remaining 20\% of data. The classifier achieved an accuracy of \textbf{94.6\%}, where 95\% of F1-score for "with control" class, and 94\% of F1-score for "without control" class\footnote{The implementation steps and results in Python can be downloaded from:~\url{https://github.com/rezatavakoli/ICALT2020_metadata}}. Moreover, we extracted the importance value of each feature for the classification task. Table~\ref{tbl-features} represents the features of our model and their importance score [0-1]. 
The importance values reveal the effect of each feature in our prediction model. The model assigns the highest value to the \textit{Availability Score} and \textit{Normal Score} features, which are the indicators we proposed. Thus, we can infer that these two indicators can illustrate the quality of OER metadata.

\begin{table}[h]
\centering
\caption{OER quality prediction model features}
\label{tbl-features}
\begin{tabular}{|p{4.3cm}|c|}
\hline
\rowcolor[HTML]{C0C0C0} 
\textbf{Feature} & \textbf{Importance score [0-1]} \\ \hline
Availability Score & 0.32 \\ \hline
Normal Score & 0.25 \\ \hline
Level Metadata Availability & 0.23 \\ \hline
Description Length & 0.10 \\ \hline
Title Length & 0.05 \\ \hline
Subjects Length & 0.05 \\ \hline
\end{tabular}
\vspace{-1em}
\end{table}

\section{Conclusion and Future Work} \label{sec-conclusion}
In this study, we collected and analysed the metadata of a large OER dataset to provide deeper insights into OER metadata quality, and proposed a scoring and a prediction model to evaluate the quality of OER metadata and, as a consequence, OER content quality. The model proposed in this short paper not only helps OER providers (e.g. repositories and authors) to revisit and think about the importance of their metadata quality, but also facilitate the quality control of OERs in general. These are essential in the light of the rapidly growing number of OERs and OER providers these days. 
Applying our model on our Skillscommons dataset indicated that it can detect OERs with quality control with the accuracy of \textbf{94.6\%}. 

We consider this study as one of the first important steps to propose intelligent models to improve OER metadata quality and OER content. As future work, we plan to further improve and validate our models by collecting more data from other OER repositories and consider more metadata features (e.g. text-based analysis of title and description). Additionally, we plan to validate our approach in other contexts, for instance by applying our scoring and prediction model to open educational videos on Youtube.

\bibliographystyle{IEEEtran}
\bibliography{main}

\begin{thebibliography}{10}
\providecommand{\url}[1]{#1}
\csname url@samestyle\endcsname
\providecommand{\newblock}{\relax}
\providecommand{\bibinfo}[2]{#2}
\providecommand{\BIBentrySTDinterwordspacing}{\spaceskip=0pt\relax}
\providecommand{\BIBentryALTinterwordstretchfactor}{4}
\providecommand{\BIBentryALTinterwordspacing}{\spaceskip=\fontdimen2\font plus
\BIBentryALTinterwordstretchfactor\fontdimen3\font minus
  \fontdimen4\font\relax}
\providecommand{\BIBforeignlanguage}[2]{{%
\expandafter\ifx\csname l@#1\endcsname\relax
\typeout{** WARNING: IEEEtran.bst: No hyphenation pattern has been}%
\typeout{** loaded for the language `#1'. Using the pattern for}%
\typeout{** the default language instead.}%
\else
\language=\csname l@#1\endcsname
\fi
#2}}
\providecommand{\BIBdecl}{\relax}
\BIBdecl

\bibitem{chicaiza2017recommendation}
J.~Chicaiza, N.~Piedra, J.~Lopez-Vargas, and E.~Tovar-Caro, ``Recommendation of
  open educational resources. an approach based on linked open data,'' in
  \emph{Global Engineering Education Conference}.\hskip 1em plus 0.5em minus
  0.4em\relax IEEE, 2017, pp. 1316--1321.

\bibitem{kiraly2018measuring}
P.~Kir{\'a}ly and M.~B{\"u}chler, ``Measuring completeness as metadata quality
  metric in europeana,'' in \emph{2018 IEEE International Conference on Big
  Data (Big Data)}.\hskip 1em plus 0.5em minus 0.4em\relax IEEE, 2018, pp.
  2711--2720.

\bibitem{ochoa2009automatic}
X.~Ochoa and E.~Duval, ``Automatic evaluation of metadata quality in digital
  repositories,'' \emph{International journal on digital libraries}, vol.~10,
  no. 2-3, pp. 67--91, 2009.

\bibitem{Tani2013}
A.~Tani, L.~Candela, and D.~Castelli, ``{Dealing with metadata quality: The
  legacy of digital library efforts},'' \emph{Information Processing and
  Management}, vol.~49, no.~6, pp. 1194--1205, 2013.

\bibitem{Trippel2014}
T.~Trippel, D.~Broeder, M.~Durco, and O.~Ohren, ``{Towards automatic quality
  assessment of component metadata},'' \emph{Proceedings of the 9th
  International Conference on Language Resources and Evaluation, LREC 2014},
  pp. 3851--3856, 2014.

\bibitem{nichols2008lightweight}
D.~M. Nichols, C.-H. Chan, D.~Bainbridge, D.~McKay, and M.~B. Twidale, ``A
  lightweight metadata quality tool,'' in \emph{Proceedings of the 8th
  ACM/IEEE-CS joint conference on Digital libraries}, 2008, pp. 385--388.

\bibitem{bruce2004continuum}
T.~R. Bruce and D.~I. Hillmann, ``The continuum of metadata quality: defining,
  expressing, exploiting,'' in \emph{Metadata in Practice}.\hskip 1em plus
  0.5em minus 0.4em\relax ALA editions, 2004.

\bibitem{Ochoa2006}
X.~Ochoa and E.~Duval, ``{Quality Metrics for Learning Object Metadata},''
  \emph{World Conference on Educational Multimedia, Hypermedia and
  Telecommunications}, no. 2004, 2006.

\bibitem{romero2019proposal}
A.~Romero-Pelaez, V.~Segarra-Faggioni, N.~Piedra, and E.~Tovar, ``A proposal of
  quality assessment of oer based on emergent technology,'' in \emph{2019 IEEE
  Global Engineering Education Conference (EDUCON)}.\hskip 1em plus 0.5em minus
  0.4em\relax IEEE, 2019, pp. 1114--1119.

\bibitem{ushakova_2015}
\BIBentryALTinterwordspacing
S.~Ushakova, ``Usability of metadata standards for open educational
  resources,'' Oct 2015. [Online]. Available:
  \href{https://hclemuseum.wordpress.com/2015/10/02/usability-of-metadata-standards-for-open-educational-resources/}{https://hclemuseum.wordpress.com/2015/10/02/usability-of-metadata...}
\BIBentrySTDinterwordspacing

\bibitem{sicilia2005complete}
M.~A. Sicilia, E.~Garcia, C.~Pag{\'e}s, J.-J. Mart{\'\i}nez, and J.~M.
  Gutierrez, ``Complete metadata records in learning object repositories: some
  evidence and requirements,'' \emph{International Journal of Learning
  Technology}, vol.~1, no.~4, pp. 411--424, 2005.

\bibitem{margaritopoulos2012quantifying}
M.~Margaritopoulos, T.~Margaritopoulos, I.~Mavridis, and A.~Manitsaris,
  ``Quantifying and measuring metadata completeness,'' \emph{Journal of the
  American Society for Information Science and Technology}, vol.~63, no.~4, pp.
  724--737, 2012.

\bibitem{pelaez2017metadata}
A.~R. Pelaez and P.~P. Alarcon, ``Metadata quality assessment metrics into ocw
  repositories,'' in \emph{Proceedings of the 2017 9th International Conference
  on Education Technology and Computers}.\hskip 1em plus 0.5em minus
  0.4em\relax ACM, 2017, pp. 253--257.

\bibitem{margaritopoulos2008conceptual}
T.~Margaritopoulos, M.~Margaritopoulos, I.~Mavridis, and A.~Manitsaris, ``A
  conceptual framework for metadata quality assessment.'' in \emph{Dublin Core
  Conference}, 2008, pp. 104--113.

\bibitem{romero2018exploring}
A.~Romero-Pelaez, V.~Segarra-Faggioni, and P.~P. Alarcon, ``Exploring the
  provenance and accuracy as metadata quality metrics in assessment resources
  of ocw repositories,'' in \emph{Proceedings of the 10th International
  Conference on Education Technology and Computers}.\hskip 1em plus 0.5em minus
  0.4em\relax ACM, 2018, pp. 292--296.

\bibitem{gavrilis2015measuring}
D.~Gavrilis, D.-N. Makri, L.~Papachristopoulos, S.~Angelis, K.~Kravvaritis,
  C.~Papatheodorou, and P.~Constantopoulos, ``Measuring quality in metadata
  repositories,'' in \emph{International Conference on Theory and Practice of
  Digital Libraries}.\hskip 1em plus 0.5em minus 0.4em\relax Springer, 2015,
  pp. 56--67.

\end{thebibliography}

\end{document}